\documentstyle[10pt,epsf,epsfig,dp_delphititle,cite,mcite]{dp_delphi}
\makeindex
\def\DpPaperGroup{PH-EP}
\def\DpPaperRef{2005-050}
\def\DpDate{18 October 2005}
\def\DpAuthors{DELPHI Collaboration}
\def\DpSubmit{(Accepted by Eur. Phys. J. C)}
\def\DpTitle{{A Determination of the Centre-of-Mass Energy
at LEP2 using Radiative 2-fermion Events}} 
\def\DpComment{}
\def\DpEMail{}

\newcommand {\ssize}{\scriptsize}

\newcommand {\GeV} {\mbox{$\mathrm{GeV}$}}

\newcommand {\MeV} {\mbox{$\mathrm{MeV}$}}

\newcommand{\KK}{\mbox{KK}}

\newcommand{\chisq}{\mbox{$\chi^{2}$}}

\newcommand {\ffbar} {\mbox{$\mathrm{f\bar{f}}$}}

\newcommand {\ee} {\mbox{$\mathrm{e^{+}e^{-}}$}}

\newcommand {\mumu} {\mbox{$\mu^{+}\mu^{-}$}}
\newcommand {\mumug} {\mbox{$\mu^{+}\mu^{-}(\gamma)$}}
\newcommand {\tautau} {\mbox{$\tau^{+}\tau^{-}$}}

\newcommand {\qqbarg} {\mbox{$\mathrm{q\bar{q}}(\gamma)$}}

\newcommand {\eemmg} {\mbox{$\ee \rightarrow \mumug$}}

\newcommand {\eeqqg} {\mbox{$\ee \rightarrow \qqbarg$}}

\newcommand {\MZ} {\mbox{$\mathrm{M}_{\mbox{\ssize{Z}}}$}}

\newcommand {\MW} {\mbox{$\mathrm{M_W}$}}

\newcommand {\GZ} {\mbox{$\Gamma_{\mbox{\ssize{Z}}}$}}

\begin{document}
\makeatletter
\newcount\@tempcntc
\def\@citex[#1]#2{\if@filesw\immediate\write\@auxout{\string\citation{#2}}\fi
  \@tempcnta\z@\@tempcntb\m@ne\def\@citea{}\@cite{\@for\@citeb:=#2\do
    {\@ifundefined
       {b@\@citeb}{\@citeo\@tempcntb\m@ne\@citea\def\@citea{,}{\bf ?}\@warning
       {Citation `\@citeb' on page \thepage \space undefined}}%
    {\setbox\z@\hbox{\global\@tempcntc0\csname b@\@citeb\endcsname\relax}%
     \ifnum\@tempcntc=\z@ \@citeo\@tempcntb\m@ne
       \@citea\def\@citea{,}\hbox{\csname b@\@citeb\endcsname}%
     \else
      \advance\@tempcntb\@ne
      \ifnum\@tempcntb=\@tempcntc
      \else\advance\@tempcntb\m@ne\@citeo
      \@tempcnta\@tempcntc\@tempcntb\@tempcntc\fi\fi}}\@citeo}{#1}}
\def\@citeo{\ifnum\@tempcnta>\@tempcntb\else\@citea\def\@citea{,}%
  \ifnum\@tempcnta=\@tempcntb\the\@tempcnta\else
   {\advance\@tempcnta\@ne\ifnum\@tempcnta=\@tempcntb \else \def\@citea{--}\fi
    \advance\@tempcnta\m@ne\the\@tempcnta\@citea\the\@tempcntb}\fi\fi}
 
\makeatother
\begin{titlepage}
\pagenumbering{roman}
\CERNpreprint{\DpPaperGroup}{\DpPaperRef} 
\date{{\small\DpDate}} 
\title{\DpTitle} 
\address{\DpAuthors} 
\begin{shortabs} 
\noindent
\noindent 

Using \eemmg\ and \eeqqg\ events radiative to the Z pole,
\mbox{\scshape Delphi} has determined the centre-of-mass energy,
$\sqrt{s}$, using energy and momentum constraint methods.
The results are expressed as
deviations from the nominal \mbox{\scshape Lep} centre-of-mass energy,
measured using other techniques. The results are found to be
compatible with the 
\mbox{\scshape Lep}
Energy Working Group estimates for a combination of the 1997 to 2000 
data sets.

\end{shortabs}
\vfill
\begin{center}
\DpSubmit \ \\ 
\DpComment \ \\
\DpEMail \ \\
\end{center}
\vfill
\clearpage
\headsep 10.0pt
\addtolength{\textheight}{10mm}
\addtolength{\footskip}{-5mm}
\begingroup
%
\newcommand{\DpName}[2]{\hbox{#1$^{\ref{#2}}$},\hfill}
\newcommand{\DpNameTwo}[3]{\hbox{#1$^{\ref{#2},\ref{#3}}$},\hfill}
\newcommand{\DpNameThree}[4]{\hbox{#1$^{\ref{#2},\ref{#3},\ref{#4}}$},\hfill}
\newskip\Bigfill \Bigfill = 0pt plus 1000fill
\newcommand{\DpNameLast}[2]{\hbox{#1$^{\ref{#2}}$}\hspace{\Bigfill}}
%
\footnotesize
\noindent
\DpName{J.Abdallah}{LPNHE}
\DpName{P.Abreu}{LIP}
\DpName{W.Adam}{VIENNA}
\DpName{P.Adzic}{DEMOKRITOS}
\DpName{T.Albrecht}{KARLSRUHE}
\DpName{T.Alderweireld}{AIM}
\DpName{R.Alemany-Fernandez}{CERN}
\DpName{T.Allmendinger}{KARLSRUHE}
\DpName{P.P.Allport}{LIVERPOOL}
\DpName{U.Amaldi}{MILANO2}
% ...
\normalsize
\endgroup
\newpage
\titlefoot{Department of Physics and Astronomy, Iowa State
     University, Ames IA 50011-3160, USA
    \label{AMES}}
\titlefoot{Physics Department, Universiteit Antwerpen,
     Universiteitsplein 1, B-2610 Antwerpen, Belgium \\
     \indent~~and IIHE, ULB-VUB,
     Pleinlaan 2, B-1050 Brussels, Belgium \\
     \indent~~and Facult\'e des Sciences,
     Univ. de l'Etat Mons, Av. Maistriau 19, B-7000 Mons, Belgium
    \label{AIM}}
\titlefoot{Physics Laboratory, University of Athens, Solonos Str.
     104, GR-10680 Athens, Greece
    \label{ATHENS}}
\titlefoot{Department of Physics, University of Bergen,
     All\'egaten 55, NO-5007 Bergen, Norway
    \label{BERGEN}}
\titlefoot{Dipartimento di Fisica, Universit\`a di Bologna and INFN,
     Via Irnerio 46, IT-40126 Bologna, Italy
    \label{BOLOGNA}}
\titlefoot{Centro Brasileiro de Pesquisas F\'{\i}sicas, rua Xavier Sigaud 150,
     BR-22290 Rio de Janeiro, Brazil \\
     \indent~~and Depto. de F\'{\i}sica, Pont. Univ. Cat\'olica,
     C.P. 38071 BR-22453 Rio de Janeiro, Brazil \\
     \indent~~and Inst. de F\'{\i}sica, Univ. Estadual do Rio de Janeiro,
     rua S\~{a}o Francisco Xavier 524, Rio de Janeiro, Brazil
    \label{BRASIL}}
\titlefoot{Coll\`ege de France, Lab. de Physique Corpusculaire, IN2P3-CNRS,
     FR-75231 Paris Cedex 05, France
    \label{CDF}}
\titlefoot{CERN, CH-1211 Geneva 23, Switzerland
    \label{CERN}}
\titlefoot{Institut de Recherches Subatomiques, IN2P3 - CNRS/ULP - BP20,
     FR-67037 Strasbourg Cedex, France
    \label{CRN}}
\titlefoot{Now at DESY-Zeuthen, Platanenallee 6, D-15735 Zeuthen, Germany
    \label{DESY}}
\titlefoot{Institute of Nuclear Physics, N.C.S.R. Demokritos,
     P.O. Box 60228, GR-15310 Athens, Greece
    \label{DEMOKRITOS}}
\titlefoot{FZU, Inst. of Phys. of the C.A.S. High Energy Physics Division,
     Na Slovance 2, CZ-180 40, Praha 8, Czech Republic
    \label{FZU}}
\titlefoot{Dipartimento di Fisica, Universit\`a di Genova and INFN,
     Via Dodecaneso 33, IT-16146 Genova, Italy
    \label{GENOVA}}
\titlefoot{Institut des Sciences Nucl\'eaires, IN2P3-CNRS, Universit\'e
     de Grenoble 1, FR-38026 Grenoble Cedex, France
    \label{GRENOBLE}}
\titlefoot{Helsinki Institute of Physics and Department of Physical Sciences,
     P.O. Box 64, FIN-00014 University of Helsinki, 
     \indent~~Finland
    \label{HELSINKI}}
\titlefoot{Joint Institute for Nuclear Research, Dubna, Head Post
     Office, P.O. Box 79, RU-101 000 Moscow, Russian Federation
    \label{JINR}}
\titlefoot{Institut f\"ur Experimentelle Kernphysik,
     Universit\"at Karlsruhe, Postfach 6980, DE-76128 Karlsruhe,
     Germany
    \label{KARLSRUHE}}
\titlefoot{Institute of Nuclear Physics PAN,Ul. Radzikowskiego 152,
     PL-31142 Krakow, Poland
    \label{KRAKOW1}}
\titlefoot{Faculty of Physics and Nuclear Techniques, University of Mining
     and Metallurgy, PL-30055 Krakow, Poland
    \label{KRAKOW2}}
\titlefoot{Universit\'e de Paris-Sud, Lab. de l'Acc\'el\'erateur
     Lin\'eaire, IN2P3-CNRS, B\^{a}t. 200, FR-91405 Orsay Cedex, France
    \label{LAL}}
\titlefoot{School of Physics and Chemistry, University of Lancaster,
     Lancaster LA1 4YB, UK
    \label{LANCASTER}}
\titlefoot{LIP, IST, FCUL - Av. Elias Garcia, 14-$1^{o}$,
     PT-1000 Lisboa Codex, Portugal
    \label{LIP}}
\titlefoot{Department of Physics, University of Liverpool, P.O.
     Box 147, Liverpool L69 3BX, UK
    \label{LIVERPOOL}}
\titlefoot{Dept. of Physics and Astronomy, Kelvin Building,
     University of Glasgow, Glasgow G12 8QQ
    \label{GLASGOW}}
\titlefoot{LPNHE, IN2P3-CNRS, Univ.~Paris VI et VII, Tour 33 (RdC),
     4 place Jussieu, FR-75252 Paris Cedex 05, France
    \label{LPNHE}}
\titlefoot{Department of Physics, University of Lund,
     S\"olvegatan 14, SE-223 63 Lund, Sweden
    \label{LUND}}
\titlefoot{Universit\'e Claude Bernard de Lyon, IPNL, IN2P3-CNRS,
     FR-69622 Villeurbanne Cedex, France
    \label{LYON}}
\titlefoot{Dipartimento di Fisica, Universit\`a di Milano and INFN-MILANO,
     Via Celoria 16, IT-20133 Milan, Italy
    \label{MILANO}}
\titlefoot{Dipartimento di Fisica, Univ. di Milano-Bicocca and
     INFN-MILANO, Piazza della Scienza 2, IT-20126 Milan, Italy
    \label{MILANO2}}
\titlefoot{IPNP of MFF, Charles Univ., Areal MFF,
     V Holesovickach 2, CZ-180 00, Praha 8, Czech Republic
    \label{NC}}
\titlefoot{NIKHEF, Postbus 41882, NL-1009 DB
     Amsterdam, The Netherlands
    \label{NIKHEF}}
\titlefoot{National Technical University, Physics Department,
     Zografou Campus, GR-15773 Athens, Greece
    \label{NTU-ATHENS}}
\titlefoot{Physics Department, University of Oslo, Blindern,
     NO-0316 Oslo, Norway
    \label{OSLO}}
\titlefoot{Dpto. Fisica, Univ. Oviedo, Avda. Calvo Sotelo
     s/n, ES-33007 Oviedo, Spain
    \label{OVIEDO}}
\titlefoot{Department of Physics, University of Oxford,
     Keble Road, Oxford OX1 3RH, UK
    \label{OXFORD}}
\titlefoot{Dipartimento di Fisica, Universit\`a di Padova and
     INFN, Via Marzolo 8, IT-35131 Padua, Italy
    \label{PADOVA}}
\titlefoot{Rutherford Appleton Laboratory, Chilton, Didcot
     OX11 OQX, UK
    \label{RAL}}
\titlefoot{Dipartimento di Fisica, Universit\`a di Roma II and
     INFN, Tor Vergata, IT-00173 Rome, Italy
    \label{ROMA2}}
\titlefoot{Dipartimento di Fisica, Universit\`a di Roma III and
     INFN, Via della Vasca Navale 84, IT-00146 Rome, Italy
    \label{ROMA3}}
\titlefoot{DAPNIA/Service de Physique des Particules,
     CEA-Saclay, FR-91191 Gif-sur-Yvette Cedex, France
    \label{SACLAY}}
\titlefoot{Instituto de Fisica de Cantabria (CSIC-UC), Avda.
     los Castros s/n, ES-39006 Santander, Spain
    \label{SANTANDER}}
\titlefoot{Inst. for High Energy Physics, Serpukov
     P.O. Box 35, Protvino, (Moscow Region), Russian Federation
    \label{SERPUKHOV}}
\titlefoot{J. Stefan Institute, Jamova 39, SI-1000 Ljubljana, Slovenia
     and Laboratory for Astroparticle Physics,\\
     \indent~~Nova Gorica Polytechnic, Kostanjeviska 16a, SI-5000 Nova Gorica, Slovenia, \\
     \indent~~and Department of Physics, University of Ljubljana,
     SI-1000 Ljubljana, Slovenia
    \label{SLOVENIJA}}
\titlefoot{Fysikum, Stockholm University,
     Box 6730, SE-113 85 Stockholm, Sweden
    \label{STOCKHOLM}}
\titlefoot{Dipartimento di Fisica Sperimentale, Universit\`a di
     Torino and INFN, Via P. Giuria 1, IT-10125 Turin, Italy
    \label{TORINO}}
\titlefoot{INFN,Sezione di Torino and Dipartimento di Fisica Teorica,
     Universit\`a di Torino, Via Giuria 1,
     IT-10125 Turin, Italy
    \label{TORINOTH}}
\titlefoot{Dipartimento di Fisica, Universit\`a di Trieste and
     INFN, Via A. Valerio 2, IT-34127 Trieste, Italy \\
     \indent~~and Istituto di Fisica, Universit\`a di Udine,
     IT-33100 Udine, Italy
    \label{TU}}
\titlefoot{Univ. Federal do Rio de Janeiro, C.P. 68528
     Cidade Univ., Ilha do Fund\~ao
     BR-21945-970 Rio de Janeiro, Brazil
    \label{UFRJ}}
\titlefoot{Department of Radiation Sciences, University of
     Uppsala, P.O. Box 535, SE-751 21 Uppsala, Sweden
    \label{UPPSALA}}
\titlefoot{IFIC, Valencia-CSIC, and D.F.A.M.N., U. de Valencia,
     Avda. Dr. Moliner 50, ES-46100 Burjassot (Valencia), Spain
    \label{VALENCIA}}
\titlefoot{Institut f\"ur Hochenergiephysik, \"Osterr. Akad.
     d. Wissensch., Nikolsdorfergasse 18, AT-1050 Vienna, Austria
    \label{VIENNA}}
\titlefoot{Inst. Nuclear Studies and University of Warsaw, Ul.
     Hoza 69, PL-00681 Warsaw, Poland
    \label{WARSZAWA}}
\titlefoot{Fachbereich Physik, University of Wuppertal, Postfach
     100 127, DE-42097 Wuppertal, Germany \\
\noindent
{$^\dagger$~deceased}
    \label{WUPPERTAL}}
\addtolength{\textheight}{-10mm}
\addtolength{\footskip}{5mm}
\clearpage
\headsep 30.0pt
\end{titlepage}
\pagenumbering{arabic} 
\setcounter{footnote}{0} %
\large
\newcommand {\Ecm} {E$_{\mathrm{CM}}$}
\newcommand {\Ecmmm} {\mathrm{E}_{\mathrm{CM}}}
\newcommand {\DEcm} {${\Delta\mathrm{E}_{\mathrm{CM}}}$}
\newcommand {\DEcmmm} {{\Delta\mathrm{E}_{\mathrm{CM}}}}
\newcommand {\Ebeam} {E$_{\mathrm{beam}}$}
\newcommand {\Ebeammm} {$E$_{\mathrm{beam}}}
\newcommand {\DEbeam} {${\Delta\mathrm{E}_{\mathrm{beam}}}$}
\newcommand {\DEbeammm} {{\Delta\mathrm{E}_{\mathrm{beam}}}}
\newcommand {\fbar} {${\bar f}$}
\newcommand {\qbar} {\bar{q}}
\newcommand{\ALEPH}{\mbox{\scshape Aleph}}
\newcommand{\DELPHI}{\mbox{\scshape Delphi}}
\newcommand{\OPAL}{\mbox{\scshape Opal}}
\newcommand{\LTHREE}{\mbox{\scshape L3}}
\newcommand{\LEP}{\mbox{\scshape Lep}}
\newcommand{\LEPII}{\mbox{\scshape Lep2}}
\newcommand{\LEPI}{\mbox{\scshape Lep1}}
\newcommand{\Angles}{\mbox{\scshape Angles}}
\newcommand{\DELANA}{\mbox{\ttfamily DELANA}}
\newcommand{\DELSIM}{\mbox{\ttfamily DELSIM}}
\newcommand{\TANAGRA}{\mbox{\ttfamily TANAGRA}}
\newcommand{\ZEBRA}{\mbox{\ttfamily ZEBRA}}
\newcommand{\sqs} {$\sqrt{s}$}
\newcommand{\sqsmm} {\sqrt{s}}
\newcommand{\Zo} {Z$^0$}

\section{Introduction}
\label{sec:intro}

One of the main goals of \LEPII\ is the precise measurement of the
mass of the W-boson, \MW.  The aim, with the integrated luminosity
collected, is a total experimental uncertainty of about
40~\MeV$/c^2$. One important contribution to the systematic
uncertainty on \MW\ comes from the accuracy with which the
centre-of-mass energy, \Ecm, can be measured.
To match other statistical and systematic errors, the error on \Ecm\
at \LEPII\ should be below 30~\MeV, equivalent to an error on the beam
energy of less than 15~\MeV.
At \LEPI, very accurate determinations of \Ecm\ were
made which relied on precise beam energy (\Ebeam) 
measurements provided by the resonant depolarisation
technique~\cite{nrg1}. At \LEPII\ energies resonant
depolarisation is no longer possible.   Therefore a method using 
a magnetic extrapolation is employed, in which resonant
depolarisation is used at beam energies of 44-61~\GeV\
to calibrate NMR probes in selected dipole magnets.
These probes are then used to estimate the beam energy at 
\LEPII\ physics energies by extrapolating from the
lower calibration energies\cite{nrg}.

The uncertainty on the extrapolation 
procedure dominates the energy determination.
Several methods have been developed by the \LEP\ Energy Working Group
to constrain this uncertainty~\cite{nrg}.
These include the measurement of the beam deflection in a spectrometer,  
an analysis based on the evolution of the synchrotron tune with the RF
voltage,  and a comparison of the bending field as
measured by the NMRs with that sensed in the ring as a whole
by the flux loop.   The assigned error on \Ecm\ 
from combining the results
of these analyses is 20-25~MeV for the majority of \LEPII\ operation,
with a larger error of about 40~MeV in 2000,  where in addition
to the magnetic extrapolation itself,  other uncertainties are
significant~\cite{nrg}.

The collision energy can also be measured
in both hadronic and leptonic channels using the ``radiative
return'' events. In these events the effective \ee\ centre-of-mass
energy is reduced to that of an on-shell Z-boson by initial-state
radiation (ISR). The \ALEPH\ collaboration performed a measurement based
upon the latter method using just the 183~\GeV\
data\cite{aleph}. Subsequently, measurements at this and higher
energies have been carried out by the \LTHREE~\cite{L3} and
\OPAL~\cite{opal} collaborations.
This note describes a method that uses \eemmg\ and \eeqqg\ radiative return 
events to measure the difference between the centre-of-mass energy determined
 from radiative events and the value from the 
\LEP\ Energy Working Group \DEcm.  This method, particularly if results from
the four LEP experiments can be combined, provides a cross-check
against large systematic effects in the \LEP\ Energy Working Group
determination. The dimuon channel is the easiest 
in which to study the radiative return peak due to its simple
two-prong topology. However, the statistical power is much smaller
than that available from hadronic radiative returns.

The analysis used the data of the four last years of \LEP\ data taking
with the collision energy ranging from 183 to 207\,GeV. The total data
sample was organized in 8 energy points: 183, 189, 192, 196, 200, 202,
205, 207\,GeV. The measurement was performed at each energy point
separately and at the end a combined result was produced.  Details of
the components of the \DELPHI\ detector and their performances can be
found in~\cite{DELPHIDET1,DELPHIDET2}. Data from the period after 1st
September 2000, when part of the TPC, the main tracking device of
\DELPHI, had high voltage problems, were not included.

Simulation of signal and all relevant backgrounds was available for
each energy.
The \eemmg\ and \eeqqg\ events were simulated with the 
\KK\ 4.14 generator~\cite{kk}. Most of the four-fermion final states were
produced with the program described in~\cite{delphi4fgen}, based on
the WPHACT 2.0 generator~\cite{WPHACT}.
However, as discussed in~\cite{delphi4fgen}, for the 
final states {\mbox{$\ee \ffbar$}}, collisions with either
the {\mbox{$\mathrm{e^{+}}$}} or  {\mbox{$\mathrm{e^{-}}$}} emitted at less 
than $2^{\circ}$ to the beam, and
with the invariant mass of the $\ffbar$ system less than
40~GeV/$c^2$ were generated with PYTHIA 6.143~\cite{PYTHIA}.
Hadronisation was performed with PYTHIA
6.156. In addition, ARIADNE 4.08~\cite{ARIADNE} was used for
fragmentation studies. The generated events were passed through the
\DELPHI\ detector simulation program \DELSIM~\cite{DELPHIDET2}, and then 
through the same
reconstruction and analysis programs as the data.

Section~2 describes the analysis of the \eemmg\ channel, 
and Section~3 that of the \eeqqg\ channel. 
Conclusions are given in Section~4.


\section{The \eemmg\ radiative return events}

\subsection{The determination of  $\sqrt{s^{\prime}}$}
\label{sec:method}
For a centre-of-mass energy, $\sqrt{s}$, above the Z resonance the
differential distribution of the invariant mass of the \mumu\ pair,
$\sqrt{s^{\prime}}$, has two distinct peaks. One is a peak at
$\sqrt{s^{\prime}}\approx\sqrt{s}$ and the other is at
$\sqrt{s^{\prime}}\approx\MZ$. The latter peak is formed by the
radiation of one or more photons from the incoming electron or
positron, reducing the effective invariant mass of the \ee\ system to
that of an on-shell Z, where the cross-section for s-channel Z
exchange is large.
 
The lineshape at the Z has been studied in great detail at \LEPI.
The measurements of \MZ\ and \GZ\ have been made to \MeV\ precision 
\cite{zmass}. Using the knowledge of the resonance, combined with that 
of QED initial-state radiation, one can hope to model the
radiative-return peak to similar precision~\cite{jadach}.

If a variable $x=\sqrt{s^{\prime}/s}$ is defined, a distribution of
$x$ will have some value, $x_{\mathrm{Z}}$, at the position of the
radiative-return peak.  In a distribution of $\sqrt{s^{\prime}}$, the
position of the radiative return peak is always at some value
($\sqrt{s^{\prime}_{\mathrm{Z}}}\simeq \MZ$), independent of the
initial centre-of-mass energy.  Therefore, if
$\sqrt{s^{\prime}_{\mathrm{Z}}}$ is known, a measurement of
$x_{\mathrm{Z}}$ can be used to calculate \sqs\ from the expression:
\begin{center}
$\sqsmm~=~\frac{\sqrt{s^{\prime}_{\mathrm{Z}}}}{ x_{\mathrm{Z}}}$ .
\end{center}
 
However, the measurement of $x_{\mathrm{Z}}$ must be independent of $\sqrt{s}$ for this 
to be a valid method of evaluating \sqs. This is true for radiative 
\mumu\ events, which are consistent 
with only one hard ISR photon being radiated collinear to the beams. 

 For this simple case of planar events $s^\prime$ can be expressed 
in terms of only $s$ and the energy of the radiated photon, 
$E_{\gamma}$:
\begin{equation}
 s^\prime~=~s~-~2E_{\gamma} \sqrt{s}.
\label{eq:sp}
\end{equation}
 Using 4-momentum conservation, $E_{\gamma}$ can be written in terms
of the polar angles (defined relative to the incoming electron beam)
of the muons and \sqs\ alone:
\begin{equation}
 E_{\gamma}~=~\frac{|\sin(\theta_{\mu^+}+\theta_{\mu^-})|\sqrt{s}}
{\sin(\theta_{\mu^+})~+~\sin(\theta_{\mu^-})~+~|\sin(\theta_{\mu^+} + \theta_{\mu^-})|} ,
\label{eqn:eg}
\end{equation}
where 
$\theta_{\mu^-}$ and $\theta_{\mu^+}$ are the polar angles of the momentum 
vectors of the muon and anti-muon, respectively.
Substituting for $E_{\gamma}$ in equation~(\ref{eq:sp}), leads to an 
expression for $x$:
\begin{eqnarray}
\sqrt{\frac{s'}{s}}~=~x & = &
~\sqrt{\frac{\sin(\theta_{\mu^+})~+~\sin(\theta_{\mu^-})~-
~|\sin(\theta_{\mu^+}+\theta_{\mu^-})|}
{\sin(\theta_{\mu^+})~+~\sin(\theta_{\mu^-})~+
~|\sin(\theta_{\mu^+}+\theta_{\mu^-})|}}~,
\end{eqnarray}
in terms of only $\theta_{\mu^+}$ and $\theta_{\mu^-}$, as desired.

To investigate the possibility of a systematic discrepancy 
between \sqs\ from the 
method and that from the standard \LEP\ measurement,  
it is also useful to be able to combine data from many years. 
A quantity is defined that expresses the difference between the 
centre-of-mass energy by the method of using radiative-return events 
and the value from \LEP, 
\begin{equation}
\DEcmmm = \sqsmm_{\Angles} - \sqsmm_{\LEP} = \frac{\MZ}{x} -  \sqsmm_{\LEP}.
\end{equation}

If the two methods agree, this should give a distribution, centred
around zero, that takes the form of the radiative-return peak. A fit
can be performed to this distribution and a central value, giving the
overall difference, extracted.

%
\subsection{Event selection}
\label{sec:muon}
Initially the standard selection for dimuon events, as used in the
cross-section and asymmetry measurements, was applied.  A detailed
description of the selection can be found in~\cite{john2}.
To reject dimuon events which are not consistent with a single ISR
photon down the beam pipe, further
cuts were made:
\begin{itemize}
\item
 it was required that no photons were seen within the electromagnetic
 calorimetry. For this study the calorimetry includes the low-angle
 luminosity detectors, STIC, as well as the barrel and forward
 regions, collectively called ECAL. Photons of energy greater than
 500~MeV in the STIC, and greater than 2~GeV in the ECAL
 were considered. This cut removed events with a high transverse-energy
 photon;

\item
 the acoplanarity\footnote{The acoplanarity is defined as the complement of 
the angle, in the plane transverse to the beam, between the two tracks.} of 
the two muons with respect to the beam axis was
required to be less than $0.1$ rad.;

\item
 the ratio of the absolute value of the missing longitudinal momentum
(i.e. net momentum of the two muons along the electron beam direction)
to the missing energy (i.e. the nominal \LEP2\ centre-of-mass energy
minus the sum of the two muon energies) in the event had to be greater
than 0.8. This selection cut is effective against multiple ISR photons.

\end{itemize}
The latter two selections also reduce backgrounds from \tautau , four-fermion 
events and from cosmic rays. 
This leaves a background level of approximately $2\%$ in the data set used 
in this study.

\subsection{Fitting the radiative return peak}
\label{sec:fit}
%
The selected events were fitted to the expected shape 
of the differential cross-section in \DEcm\ using an unbinned maximum 
log-likelihood fit. The function, $f(x)$, used is: 
\begin{equation}
f(x)~=~\frac{{\mathcal{N}}x^4}{(x^2~-~x_{\mathrm{Z}}^2)^2 + (\GZ^{\mathrm{FIT}}x/\MZ)^2} 
\cdot(1 + {\mathcal{B}}_1/x + {\mathcal{B}}_2/x^2) ,
\label{eq:bw}
\end{equation}
where $\mathcal{N}$, $x_{\mathrm{Z}}$, $\GZ^{\mathrm{FIT}}$, ${\mathcal{B}}_1$ and 
${\mathcal{B}}_2$ are parameters determined by fitting.
This is an empirical function which fits well the shape of the
simulated data. The package {\tt MINUIT}~\cite{min} was used to maximise 
the likelihood function.   

%
Events generated by \KK~\cite{kk}\ and passed through the 
\DELPHI\ detector simulation package
were used to calibrate the fit.  It was found that the central fitted
value, calculated from the parameter $x_{\mathrm{Z}}$, did not correspond to 0,
but was systematically shifted to lower values of \DEcm, {\it i.e.}
lower values of \sqs. A number of different samples at several
centre-of-mass energies were used to evaluate the size of this bias.
The bias was found to be independent of energy over the 
range of energies considered.
Thus the bias from the combination of all the energies is used as the 
correction to the final fitted central value from data.
 
\begin{figure}[bth]
\begin{center}
  \mbox{\epsfig{file=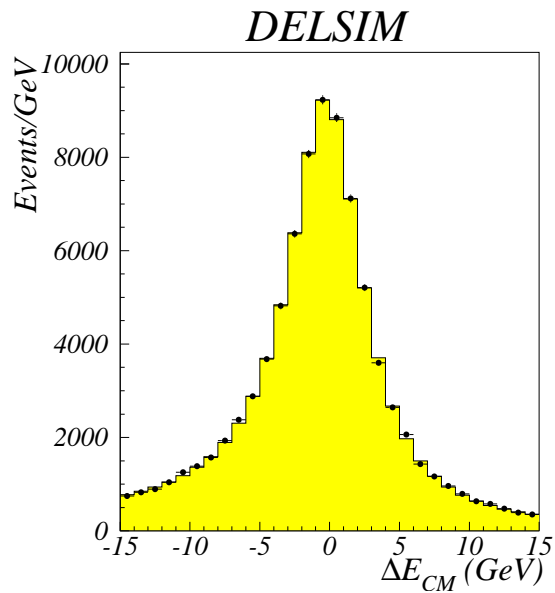,width=0.5\textwidth}} 
\caption[Fit to \DELSIM\ sample of \eemmg.]
{A fit to the \DEcm\ distribution 
(see Section~\ref{sec:method}) of fully-simulated \eemmg\ events 
using the KK generator~\cite{kk}, 
corresponding to an effective luminosity of about 86~fb$^{-1}$.  
The points show the simulated data, 
and the shaded histogram is the fitted function.}
\label{fig:kzdel}
\end{center}
\end{figure}

The function (5) was fitted to \DELSIM\ samples for \KK\ events generated at 
centre-of-mass energies at or close to those of the data. The result
of the fit to all simulated data is shown in Figure~\ref{fig:kzdel}.
The bias from this fit was found to be 
$\DEcmmm = -0.057 \pm0.022\;(stat.)$~\GeV\ 
(\chisq/N.D.o.F. = 26.8/25).

%

\subsection{Results}
\label{sec:res}
\begin{figure}[bth]
\begin{center}
\psfig{file=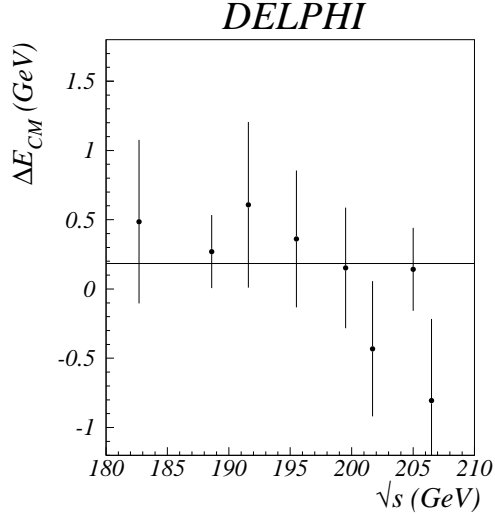,width=7cm}
\caption[]{Measured centre-of-mass energy shift at each energy point from 
\eemmg\ events. The error bars show the size of the
statistical error. The line represents the fit to all energies combined.}
\label{fig:edifvse}
\end{center}
\end{figure}

\begin{figure}[bth]
\begin{center}
  \mbox{\epsfig{file=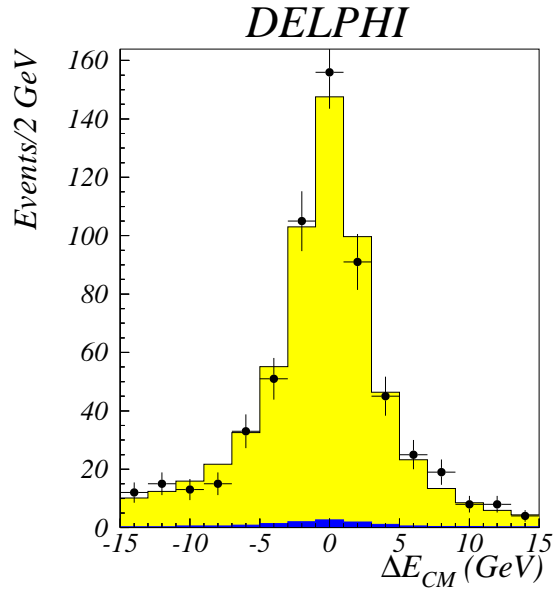,width=0.5\textwidth}}  
\caption[1997--1999 data with fitted curves for \eemmg.] 
{\DEcm\ value for 
 data (points) corresponding to the combined sample of \eemmg\ events
 collected from 1997--2000. The fitted function is shown by the light
 grey
 histogram. The dark grey 
 histogram shows the expected background. 
 The integrated luminosity is 594~pb$^{-1}$.}
\label{fig:dat}
\end{center}
\end{figure}

From the 1997--2000 data sets 600 events were selected and the \DEcm\
distribution was fitted with equation~(\ref{eq:bw}), with the
parameters ${\mathcal{B}}_1$ and ${\mathcal{B}}_2$ fixed at their
values obtained from simulation. The fit results at each energy point
are shown in Figure~\ref{fig:edifvse}. Since, within the large statistical 
errors, there is no apparent dependence on energy, the 1997-2000 data sets 
were combined. The fit is shown in Figure~\ref{fig:dat}.
The result of the fit was:
\begin{eqnarray*}
 \DEcmmm\ = +0.184 \pm 0.150 \;(stat.)\;\GeV \;(\chisq/\mbox{N.D.o.F.} = 8.6/12).
\end{eqnarray*} 

\noindent
The statistical error was verified using the ``bootstrap'' 
method~\cite{bootstrap}. In
this method, random subsamples of simulated events of the same size as the
real sample size are selected and for each of them the result is
computed in the same way as for the real sample. This gives some
distribution of the resulting value.  The spread of this distribution
was compatible with the statistical error assigned by {\tt MINUIT}
(i.e. the ``pull value'' was found to be $1.01$). 
However the average statistical
uncertainty on the centre-of-mass energy of the samples was found to be
0.172~$\GeV$, somewhat larger than that found with the real sample.

\subsection{Systematics}
\label{sec:syst}
The systematic uncertainties are presented below. They include contributions 
from the modelling of the ISR spectrum as well as effects expected 
from reconstruction and from the \LEP\ beam parameters.

\subsubsection{Systematics from the bias}
\label{sec:biassys}
The bias estimation has a systematic uncertainty on \DEcm\ of $\pm 22$~\MeV, 
due to the limited statistics of fully-simulated events.

\subsubsection{Systematics from radiative photonic correction modelling}
\label{sec:isrsyst}
The effects of the QED modelling were studied at
generator level using the \KK~\cite{kk} program. 
The \KK\ generator calculates photon emission from initial beams and 
outgoing fermions to second order, and includes 
ISR$\otimes$FSR interference also to this order.

To make some estimate of the magnitude of the effect of imperfections
in the radiative photon modelling on the bias, samples of 500,000 \KK\
events were generated at each of the centre-of mass energies used and
fitted with the function given in equation~(5). Following the advice
given by the authors of \KK~\cite{kk}, an estimate of the uncertainty of the
ISR modelling was obtained by weighting the events generated with the
default CEEX2 (${\cal O}(\alpha^2)$) approximation to correspond to
the CEEX1 (${\cal O}(\alpha)$) approximation, and fitting equation~(5)
to the weighted events. Half the difference in the resulting bias was
taken as the systematic uncertainty due to this effect. It amounted to
$2.5\pm0.7$~\MeV\ on \DEcm.

\subsubsection{Mismeasurement of the DELPHI aspect ratio}
\label{sec:aspect}
The aspect ratio is defined as the ratio of the length to the 
width of the detector. It is limited to the precision to which 
the position and dimensions of the Vertex Detector (VD) can be measured. 
The effect of a mismeasurement of the aspect ratio is to introduce a
bias on the measurement of the polar angle, $\theta$. A positive
(negative) bias on
$\tan\theta$ would systematically increase (decrease) the opening angle of the
two muons and lead to an underestimate (overestimate) of $\sqrt{s}$.

The correspondence of hits in overlapping silicon modules is sensitive
to a misalignment of the Vertex Detector. In fact the study of these overlaps 
constitutes an essential part of the procedure for the alignment of the Vertex
Detector. The alignment was carried out independently for each year of 
operation.    
A careful analysis of the data of all years shows that a residual bias
of $6\times 10^{-5}$ can be excluded at the $95\%$ confidence
level. However a study of the behaviour of the \DELPHI\ silicon
modules~\cite{LEP2SIL} shows that the centre of charge of the holes (and
electrons) differs from the mid-plane of the detectors by
$10-20$~$\mu$m. In the alignment procedure this difference would be
automatically accounted for. However much less is known about the
displacement of the electrons (used in the measurement of $z$,
along the beam axis, 
and hence $\theta$) since the necessary overlaps are much fewer. Taking
$\pm10$~$\mu$m as an effective uncertainty on the location of the
sensitive region of the detectors, and allowing for the fact that the
polar angles of some tracks are determined from signals in the silicon
detectors themselves (without using the interaction point as a
constraint) it is concluded that a reasonable estimate of the aspect
ratio uncertainty is $3\times 10^{-4}$.
Such a bias would correspond to a shift of \DEcm\ of about $\pm$29~\MeV.


\subsubsection{Systematics from LEP beam parameters}
\label{sec:leppars}
In general the colliding $\mathrm{e^{+}}$ and $\mathrm{e^{-}}$ are not
collinear at the interaction point, and their energies are not
equal. This leads to a boost of the centre-of-mass system relative to
\DELPHI\ and in an individual event will lead to a change in the fitted
centre-of-mass energy. However, averaged over all event orientations,
the effects are much reduced. Similarly, inequality of the beam
energies or a tilt of the beams with respect to the \DELPHI\
coordinate system will lead to errors in the fitted energies. Studies
at generator level with beam acollinearities and tilts up to $500~\mu$rad and beam 
energy differences up to 100~\MeV, have shown that the overall effects would be 
approximately 1~\MeV. However the spread in energies of the electron and 
positron beams, which can give an rms spread in the centre-of-mass 
energy of $\pm 250$ MeV at high energy, results in a correction of 6~MeV.

\subsubsection{Muon polar angle and momentum reconstruction}
\label{thetasmear}
Studies at generator level showed that random errors in the muon polar
angle due to reconstruction inaccuracies in \DELPHI\ lead to
systematic shifts of the fitted energy. In principle, if correctly
estimated, these are taken into account in the simulation. The
selection on energy and momentum imbalance also makes use of the
measured momenta of the muons, so in principle is affected by bias or
additional smearing in this quantity.  However it was estimated that
the overall effect of these imperfections led to an uncertainty in the
centre-of-mass energy of less than 2~MeV.

\subsubsection{Backgrounds}
Backgrounds from four-fermion and $\tau$-pair production have been
included in the simulated samples from which the bias was
calculated. Inclusion of the 2$\%$ estimated backgrounds changed the
bias by $0.4~\MeV$ on \DEcm, with a statistical uncertainty of $\pm 4~\MeV$, 
due to the size of the samples of simulated background events. 
The latter value was taken as the systematic uncertainty due to the background.
The energies of the simulated backgrounds were not varied during the
fit. It was estimated that this introduces a negligible additional
systematic uncertainty.

\subsubsection{Knowledge of the Z mass}
As can be seen from equation (4), there will be an uncertainty in the estimate 
of \DEcm, due to the uncertainty of the Z mass as measured at  
\LEPI~\cite{zmass}. The resulting uncertainty from this source amounts to $\pm 4~\MeV$.

\subsubsection{Summary of systematic uncertainties}
\label{sec:systab}
Table~\ref{tab:sysmm} gives a breakdown of the systematics on \Ecm. 
The total systematic uncertainty on \DEcm\ is 38~\MeV. Thus the dimuon
events give a value of:
\begin{eqnarray*}
\DEcmmm = +0.241 \pm0.150 \;(stat.) \pm0.038 \;(syst.)~\mbox{\rm GeV}.
\end{eqnarray*}
%
\begin{table}[h]
\begin{center}
\begin{tabular}{|l||c|} \hline 
 Source & Error on \DEcm ~(\MeV) \\ \hline \hline
{ Bias} & ~22 \\
{ QED modelling}  & ~~3 \\ 
{ Aspect ratio} & ~29 \\
{ \LEP\ parameters} & ~~6 \\
{ Angular resolution} & ~~2 \\
{ Backgrounds} & ~~4 \\
{ Z mass} & ~~4 \\ 
\hline \hline
{Total systematic error} & ~38 \\ \hline
\hline
{Statistical error} & 150 \\ \hline
\end{tabular}
\caption[Summary of systematic  errors]            
{The sources and associated values of uncertainties on \DEcm, 
measured from radiative dimuon events.}
\label{tab:sysmm}
 \end{center}
\end{table}

\section{The \eeqqg\ radiative return events}

\subsection{Track and event selection}
Events of the type \eeqqg\ give rise to a large hadronic multiplicity.
The analysis was based mainly on the charged particle tracks
reconstructed in each event.  Track selection criteria were applied in
order to reject tracks originating from secondary interactions as well
as poorly measured tracks. The charged particles were required to have momentum
larger than 0.4\,GeV$/c$, but not exceeding 1.5 times the beam energy,
momentum resolution better than 100\%, and transverse and longitudinal
impact parameters less than 4~cm and 10~cm, respectively. A more
detailed description of the selection can be found in
\cite{john2}.

All hadronic events were required to have at least 7 charged particles
and a transverse energy greater than 20\% of the collision energy. The
transverse energy was calculated on the basis of the transverse
momentum of each selected track and on the electromagnetic showers
reconstructed in the HPC and the FEMC with a shower energy above
500~\MeV. No other cuts were used, in particular for suppression of the
WW and ZZ backgrounds, in order to avoid a possible bias to the
results. The selection efficiency for hadronic events was slightly
less than 80\%.

\subsection{Fitting of the radiative return peak}
%
In the $s'$ determination a two-jet configuration was forced. If no ISR
photon candidate was observed then the reduced energy was computed
from the angles of the jets assuming a photon escaped into the beam pipe.
If an isolated energetic photon was observed then the value of $s'$
was derived from the combined jet and photon directions, assuming 
that additional energy could be radiated inside the beam pipe. If no
solution was found, the reduced energy was derived from 
the measured energy of the isolated photon alone. 

\begin{figure}
\begin{center}
\epsfig{file=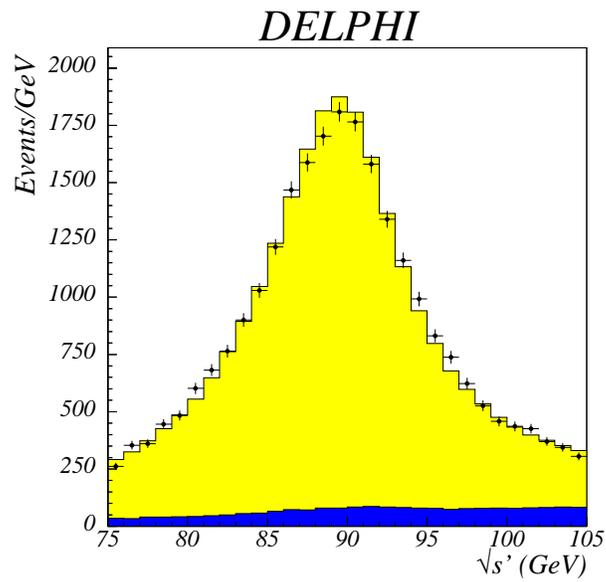,width=0.5\textwidth}\\
\caption[]{Reduced energy distribution of \eeqqg\ events. Points represent the data, light grey
area the signal simulation, dark grey 
area the background simulation. 
The integrated luminosity is approximately 600~pb$^{-1}$.
The distributions combine all energy points 183-207\,GeV.}
\label{fig:sprime}
\end{center}
\end{figure}

Figure~\ref{fig:sprime} compares $\sqrt{s'}$ distributions
between data and simulation~\cite{kk} around the radiative return peak. 
In the analysis the interval of 75-105\,GeV of the $\sqrt{s'}$
distribution was used with bins of width 1\,GeV.

\subsubsection{Method of Monte Carlo reweighting}
%
The method of this analysis was to compare the Z-boson mass, \MZ,
measured from data, to its nominal value and to find the centre-of-mass
energy shift, \DEcm, from their difference, $\Delta \MZ $, using:  
\begin{eqnarray*}
\DEcmmm = - \frac{\Delta \MZ\ }{\MZ\ } \cdot \Ecmmm .
\end{eqnarray*}

The approach used for the \MZ\
determination is based on varying the \MZ\ value in simulation and
finding the value which provides the best agreement of the $\sqrt{s'}$
distribution between data and simulation.
\MZ\ was varied by reweighting the simulation using a
Breit-Wigner-like function $f(s',\MZ)$:
\[
weight(s',\MZ^{new}) = \frac{f(s',\MZ^{new})}{f(s',\MZ^{nom})},
\]
\[
f(s',\MZ) = \frac{(1 + P \cdot (\sqrt{s'}-\MZ))\cdot s'} 
                 {(s'-{\MZ}^2)^2 + (s' \cdot \GZ/\MZ)^2},
\]
where $P$ is a skewness parameter (set to 0.01\,GeV$^{-1}$),
$\sqrt{s'}$ is the generated reduced energy in the given event,
$\MZ^{nom}$ the nominal value of $\MZ$ and $\MZ^{new}$ the new value of \MZ.
This particular form of the function $f(s',\MZ)$ was chosen because it
describes well the shape of the Z mass peak of the generated 
$\sqrt{s'}$ distribution, and this is used to set the value of $P$. 

\subsubsection{Fitting simulation to data}
%
The determination of \MZ\ was performed by fitting simulation to data
using a log-likelihood method. The log-likelihood of the real
distribution, given the simulated $\sqrt{s'}$ distribution, was
computed as:
\begin{equation}\label{eq:loglike}
{\cal L} = 2\sum_{i=1}^n(f_i - y_i\ln{f_i}),
\end{equation}
where $n$ is the number of bins in the distribution histogram,
$y_i$ and $f_i$ are the content of the $i$-th bin in the
real and simulated distributions, respectively.

For a cross-check a chi-square method was used as well.
The chi-square of the difference was computed as:
\begin{equation}\label{eq:chi2}
\chi^2 = \sum_{i=1}^n\left( \frac{f_i - y_i}{\sqrt{y_i}}\right)^2.
\end{equation}


When the simulation has been reweighted to a new \MZ\ value the
content of the simulated distribution $f_1...f_n$ was recomputed
accordingly.  By a detailed scan, the dependence of $\chi^2$ and
${\cal L}$ on \MZ\ was found to be quadratic over the region of
interest. It can be written as:
\begin{equation}  \label{eq:parabola}
\chi^2(\MZ) = {\chi_{min}}^2 + \frac{(\MZ-\MZ^{min})^2}{\sigma^2}.
\end{equation}

Three points are needed to find the minimum of the parabola (\ref{eq:parabola}).
The most convenient choice of points is $\MZ-0.5$\,GeV, \MZ\ and $\MZ+0.5$\,GeV.
After reweighting the simulation to $\MZ-0.5$\,GeV and $\MZ+0.5$\,GeV and
computing ${\cal L}$ or $\chi^2$ at all three points it is straightforward
to compute $\MZ^{min}$ as well as $\sigma$, which gives the statistical error
on $\MZ^{min}$.

\subsection{Corrections in the forward region}
\begin{figure}[bt]
\begin{center}
\epsfig{file=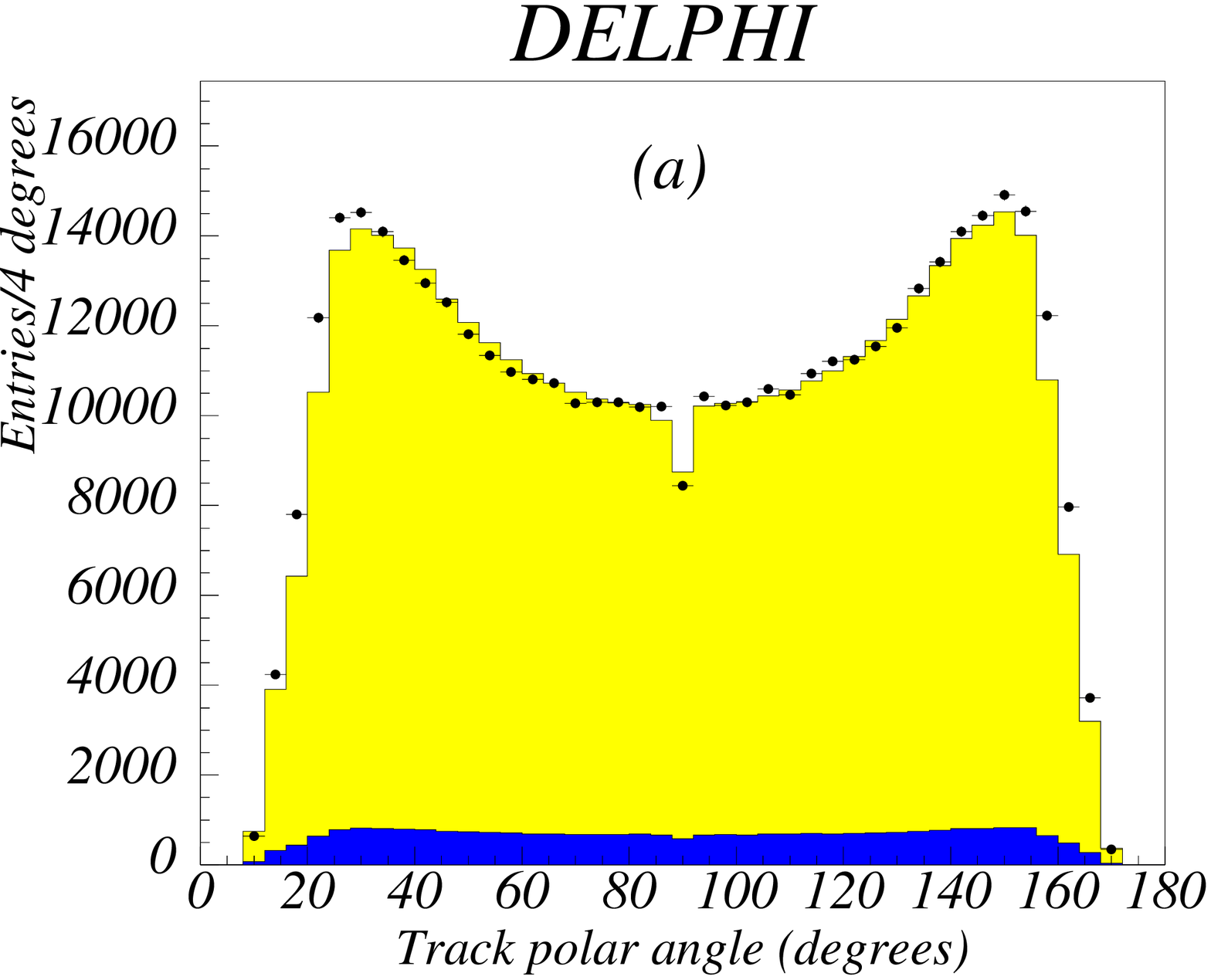,width=0.5\textwidth}\\
\vspace{-0.1cm}
\epsfig{file=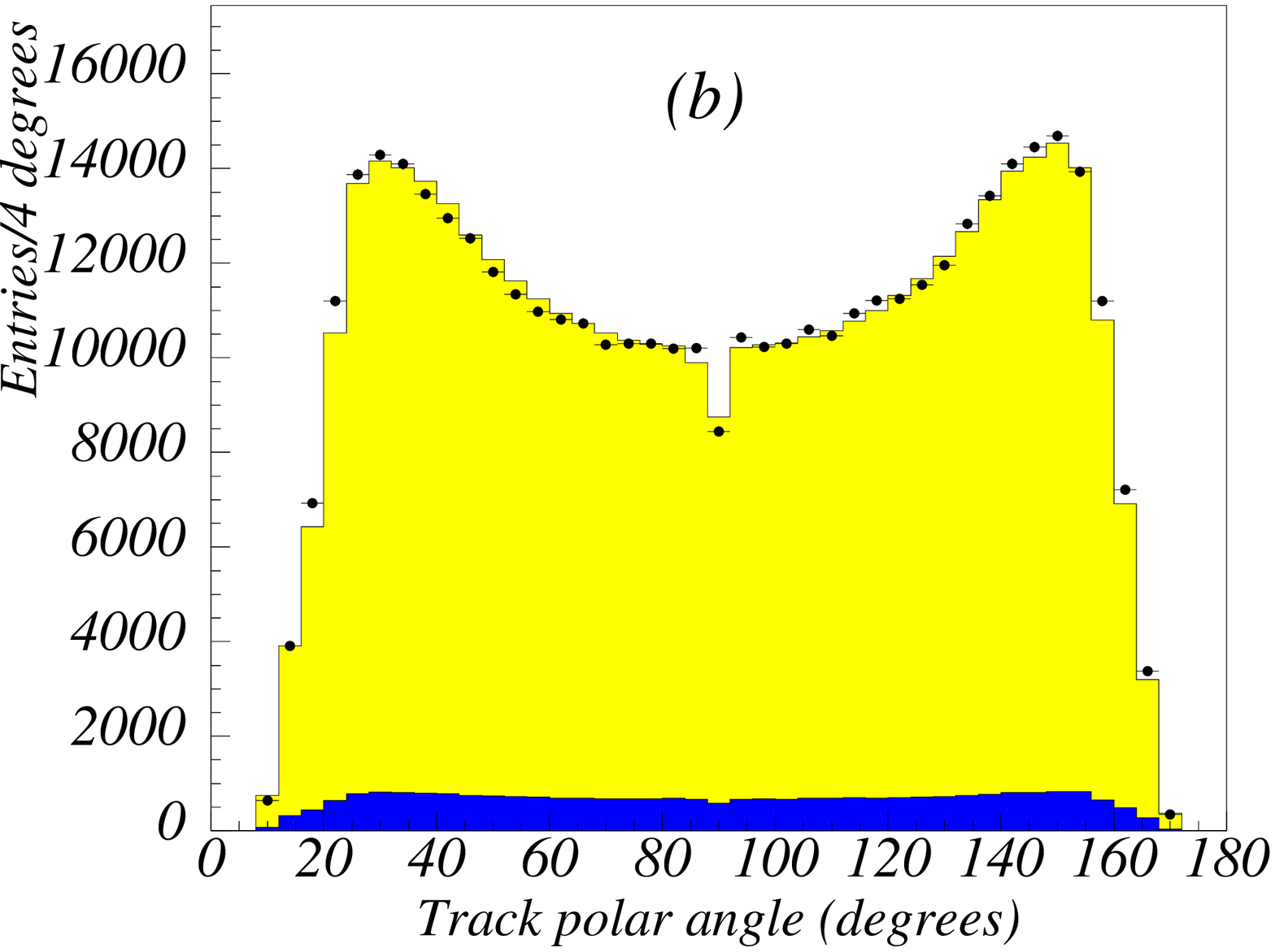,width=0.5\textwidth}\\
\caption[]{Distribution of track polar angle of radiative return \eeqqg\ events
(a) before and (b) after corrections in the forward region. 
Points represent data, light grey
area the signal simulation, dark grey 
area the background simulation. 
The distributions combine all energy points 183-207\,GeV.}
\label{fig:theta}
\end{center}
\end{figure}

One of the main problems to be dealt with in the analysis was a
noticeable excess of tracks at low polar angles (forward tracks) in
data as compared to simulation. The effect is shown in
Figure~\ref{fig:theta}a. The most likely cause of this effect is an
underestimation in the simulation of the track reconstruction
efficiency for low-momentum particles at low polar angles.


A tuning was applied to the data in order to correct this
discrepancy. It consisted in removing some fraction of forward tracks
on a random basis. A track could be removed from an event with some
probability ${\cal P}$ computed from a ratio between the number of
simulated $N_{sim}$ and real $N_{real}$ tracks:
\begin{eqnarray*}
{\cal P} = 1-\frac{N_{sim}}{N_{real}}.
\end{eqnarray*}
In computing the correction it was taken into account that the size of 
the discrepancy
depends on the track momentum and polar angle. The momentum range, as well
as polar angle range, were split into a certain number of bins.
The ratio $N_{sim}/N_{real}$ was computed in each bin and the correction
was applied accordingly.

For computation of the correction ratios $N_{sim}/N_{real}$, the data
collected in calibration runs at the Z peak were used. Such data are
available for every year of \LEPII\ data taking. They provide much
more statistical power for computation of the tuning coefficients than the
high energy data, and the corrections are uncorrelated with the
\LEPII\ beam energy.

Figure~\ref{fig:theta}b shows the polar angle distribution after the
correction is applied. Some residual difference still remains because
the size of the discrepancy in high energy and in on-peak data was
slightly different.

This correction changed significantly the results because radiative
events are boosted forward and therefore are sensitive to changes in
forward tracking.  The measured average shift in the centre-of-mass
energy changed as a result of the correction from +206\,MeV to
$-$116\,MeV.

It should be noted that this effect does not influence the muon-pair
result. The defect in the simulation would show itself as a difference
in efficiency between simulation and data. However as part of the
dimuon analysis, the detection efficency in simulation is tuned to
that measured in data. Furthermore the effect is concentrated at low
track momenta. Hence the results presented in Section 2 are not affected.

\subsection{Results}
\begin{figure}[bth]
\begin{center}
\epsfig{file=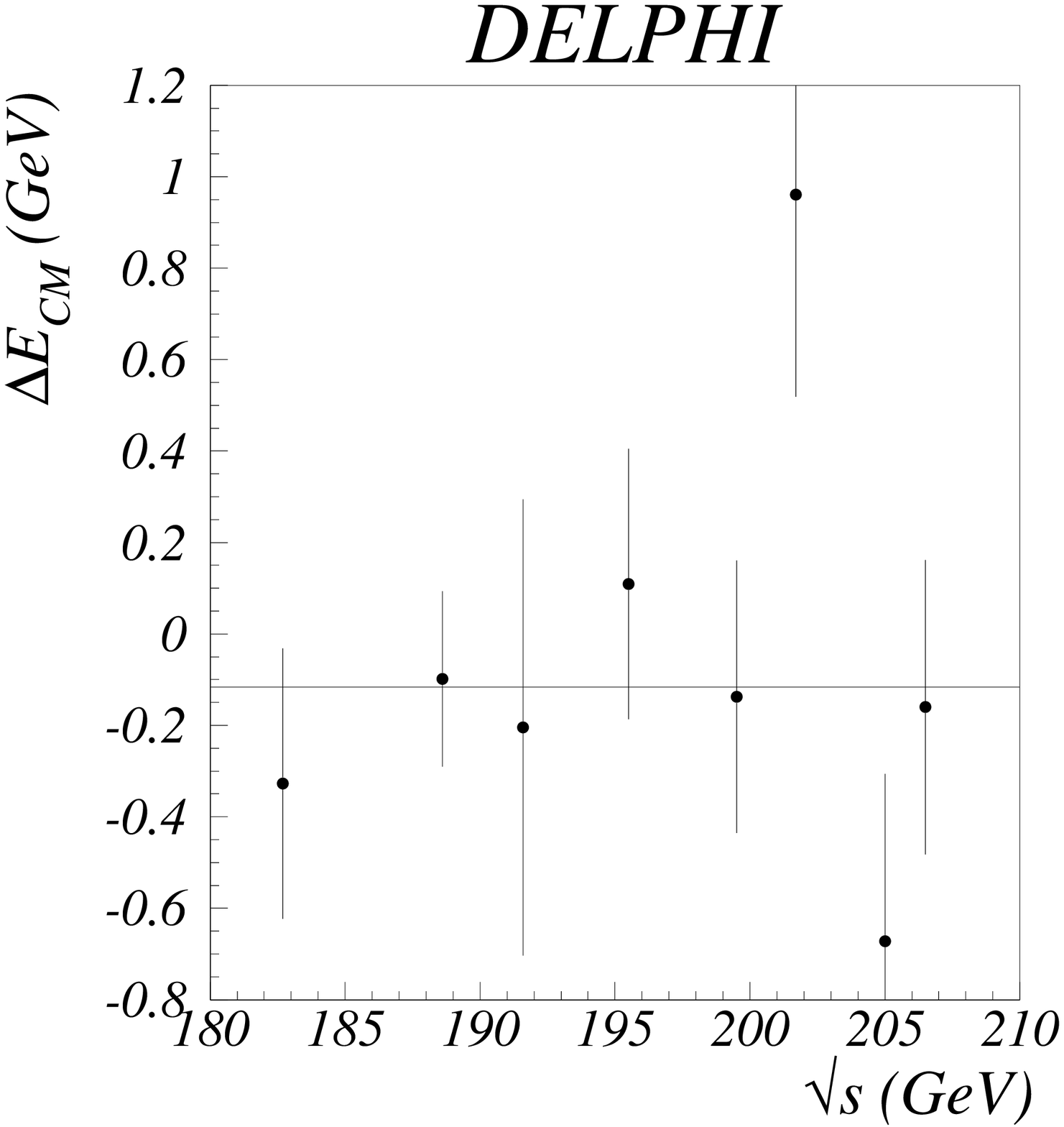,width=7cm}\\
\caption[]{Measured centre-of-mass energy shift at each energy point from 
\eeqqg\ events. The error
bars show the size of the statistical error. The line represents the
average (\chisq/N.D.o.F. = 9.4/7).}
\label{fig:fit}
\end{center}
\end{figure}

The results obtained from fitting simulation to data by the
log-likelihood method at each energy point are shown in 
Figure~\ref{fig:fit}. 
The average shift of the measured centre-of-mass energy
with respect to the \LEP\ value is found to be $-116 \pm 106 \;(stat.)$\,MeV.

%
In the analysis the statistical error on $\MZ^{min}$ is computed 
as $\sigma$ from equation (\ref{eq:parabola}).
In order to verify that it gives a correct estimate of the statistical error, 
two other methods of the statistical error determination were used.

The first one is the so-called ``Jackknife'' method~\cite{jackknife}.
It describes the measurement of the change in the result if one single
event is taken out of the sample.  If one event is removed out of the
bin $i$, the change of the result $R$ is $\Delta_i=R-R_i$. The
variance ${\sigma_R}^2$ of the result is then equal to the sum of
${\Delta_i}^2$ over all bins times the number of entries $y_i$,
${\sigma_R}^2=\sum_i{\Delta_i}^2\cdot y_i$. Thus $\sigma_R$ should
give the same error on \MZ\ as before.

The statistical error was also checked using the ``bootstrap''
method~\cite{bootstrap} described in Section~\ref{sec:res}.

Statistical errors computed with the methods described above were
found to be consistent within a few percent with the error computed
from equation~(\ref{eq:parabola}).

The contribution of the statistics of the Monte Carlo sample amounts 
to an uncertainty of $\pm 14~\MeV$ on \DEcm.

\subsection{Systematics}

\subsubsection{Uncertainty on corrections in the forward region}


The stability of the \Ecm\ determination was studied using alternative
methods for the determination of $\sqrt{s^{\prime}}$ (using also neutrals or a
kinematic fit). A correction procedure based on the high energy data
rather than calibration data at the Z peak was also tried, although
this could not be used for a result since the correction is not
independent of the assumed centre-of-mass energy. A reweighting of
tracks in data, rather than their removal, was also used.
Considering the spread of the results,
and also the size of remaining data-simulation discrepancies (for
example those visible in Figure~\ref{fig:theta}b) it was estimated
that a systematic uncertainty of $\pm$120\,MeV on \Ecm\ should be
assigned, due to the forward track correction and other imperfections
in the simulation.


\subsubsection{ISR modelling uncertainty}
To study the uncertainty related to ISR modelling in simulation,
two \eeqqg\ Monte Carlo samples,
made with different generators, were compared. One sample was
generated with \KK\ and another one with PYTHIA 6.153. 

The comparison revealed no statistically significant effect. When one
simulation was fitted with the other, the deviation of \MZ\ from the
nominal value was found to be consistent with zero. The
statistical error of this measurement was approximately 20\,MeV on \Ecm.

In addition, as for the dimuon events, studies were made at generator
level using the KK generator. Samples of 100,000 \eeqqg\ events were
generated at each of the energies used. Half of the difference between
the CEEX2 and CEEX1 approximations amounted to $0.2\pm0.4~\MeV$ on
\DEcm. However in the DELPHI simulation, events are generated with
KK without ISR$\otimes$FSR interference. The hadronisation and FSR were
generated using PYTHIA. The effect of removing ISR$\otimes$FSR in KK-generated
events amounted to a change in \DEcm\ of $4.4\pm5.4~\MeV$.

The uncertainty connected with the ISR modelling was therefore taken
as $\pm6~\MeV$ on \DEcm.

\subsubsection{Fragmentation uncertainty}
A similar approach was used for estimating the fragmentation
uncertainty.  A sample generated with \KK\ and hadronised with PYTHIA
was compared to a sample generated with \KK\ and hadronised with
ARIADNE~\cite{ARIADNE}. Both PYTHIA and ARIADNE had been consistently
tuned to the \DELPHI\ data. Fitting one simulation to the other gave a
few MeV shift in \MZ, with the statistical error on the difference of
about 10\,MeV. Therefore the fragmentation uncertainty was estimated to be
$\pm20~\MeV$ on \Ecm.

\subsubsection{Four-fermion background uncertainty}
The majority of the four-fermion background is concentrated at high
values of $s'$. Around the Z mass peak the background amounts to 8-10\%
(depending on energy point) of the signal and it has an almost flat 
distribution with $s'$ (see Figure~\ref{fig:sprime}). Thus, a bias 
in the background will affect the results only weakly.

The uncertainty due to four-fermion background was estimated by
scaling the background cross-sections by $\pm5$\%. It changed the
result by $\pm12$\,MeV.  It was concluded that a 10\,MeV error is
sufficient to cover the background uncertainty. The energies of the 
simulated backgrounds were not varied during the fit. It was estimated 
that this introduces a negligible additional systematic uncertainty.

\subsubsection{Aspect ratio uncertainty}
As discussed in Section~\ref{sec:aspect}, 
uncertainties in the aspect ratio can bias the
reconstruction of jet angles and therefore the $s'$ determination as well.

Its effect on the centre-of-mass energy measurement was checked by
introducing in simulation an artificial aspect ratio bias of a size of
$3\times 10^{-4}$. It consisted in scaling the longitudinal component
of the momenta of charged particle tracks by this amount. The resulting change
of the measured centre-of-mass energy was found to be 24\,MeV. This
value was taken as the aspect ratio uncertainty.

\subsubsection{Systematics from LEP beam parameters}
\label{sec:lepparsq}
The effects of the \LEP\ beam parameters on the energy measurement
were taken to be the same as those found in the generator studies of
dimuon events, namely an uncertainty of $\pm 6~\MeV$ on \DEcm.

\subsubsection{Knowledge of the Z mass}
As in the dimuon analysis, there will be an uncertainty in the
estimate of \DEcm\ of $\pm 4~\MeV$, due to the uncertainty of the Z
mass as measured~\cite{zmass} at \LEPI.

\subsubsection{Error summary}
Table~\ref{tab:syst} summarizes all contributions to the error
of the measurement.
The total systematic uncertainty on \DEcm\ is $126~\MeV$. 
Thus the \eeqqg\ events gave a value of:
\begin{eqnarray*}
\DEcmmm = -0.116 \pm0.106 \;(stat.) \pm0.126 \;(syst.)~\mbox{\rm GeV}.
\end{eqnarray*}

\begin{table}[h]
\begin{center}
\begin{tabular}{|l|c|}
\hline
Source & Error on \DEcm\ (\MeV) \\
\hline
\hline
Data/MC disagreement & 120  \\
ISR modelling   & ~~6  \\
Fragmentation  & ~20  \\
Background     & ~10  \\
Aspect ratio   & ~24 \\
\LEP\ parameters & ~~6 \\
MC statistics  & ~14  \\
Z mass & ~~4 \\
\hline
\hline
Total systematic error & 126  \\
\hline
\hline
Statistical error & 106  \\
\hline
\end{tabular}
\caption[]{Contributions to the uncertainty on the centre-of-mass 
energy shift, as measured from radiative hadronic events.}
\label{tab:syst}
\end{center}
\end{table}

\section{Conclusions}
\label{sec:conclusions}
High-energy data collected in \DELPHI\ during the years 1997-2000 were 
analysed in order to cross-check the \LEP\ energy measurement. The
average difference between the centre-of-mass energy measured from radiative
return \eemmg\ events and the energy provided by the \LEP\ Energy
Working Group was found to be $+0.241 \pm 0.150 \;(stat.)
\pm0.038 \;(syst.)$ GeV. The corresponding analysis using the \eeqqg\ events
yielded a difference of $-0.116 \pm 0.106 \;(stat.) \pm 0.126 \;(syst.)$ GeV.
Taking account of correlated systematic errors (aspect ratio, \LEP\
parameters, Z mass), the two results can be combined to give an
overall \DELPHI\ result:
\begin{eqnarray*}
\DEcmmm = +0.073 \pm0.094 \;(stat.) \pm0.065 \;(syst.)~\mbox{\rm GeV}.
\end{eqnarray*}

Thus the \DELPHI\ data are compatible with the values reported by
the \LEP\ Energy Working Group~\cite{nrg},
who quote values of the LEP centre-of-mass
energy with uncertainties ranging from $\pm 20$ to $\pm 40$ MeV, depending
on the year of operation.
%
\subsection*{Acknowledgements}

We would like to acknowledge the assistance of the authors of
ref.~\cite{kk} for clarification of several points connected with the
\KK\ generator.

We are greatly indebted to our technical collaborators, to former
members of the CERN-SL division for their excellent performance of the
LEP collider, and to the funding agencies for their support in building
and operating the DELPHI detector. \\
We acknowledge in particular the support of \\
Austrian Federal Ministry of Education, Science and Culture,
GZ 616.364/2-III/2a/98, \\
FNRS--FWO, Flanders Institute to encourage scientific and technological 
research in the industry (IWT), Belgium,  \\
FINEP, CNPq, CAPES, FUJB and FAPERJ, Brazil, \\
Czech Ministry of Industry and Trade, GA CR 202/99/1362,\\
Commission of the European Communities (DG XII), \\
Direction des Sciences de la Mati$\grave{\mbox{\rm e}}$re, CEA, France, \\
Bundesministerium f$\ddot{\mbox{\rm u}}$r Bildung, Wissenschaft, Forschung 
und Technologie, Germany,\\
General Secretariat for Research and Technology, Greece, \\
National Science Foundation (NWO) and Foundation for Research on Matter (FOM),
The Netherlands, \\
Norwegian Research Council,  \\
State Committee for Scientific Research, Poland, SPUB-M/CERN/PO3/DZ296/2000,
SPUB-M/CERN/PO3/DZ297/2000, 2P03B 104 19 and 2P03B 69 23(2002-2004)\\
FCT - Funda\c{c}\~ao para a Ci\^encia e Tecnologia, Portugal, \\
Vedecka grantova agentura MS SR, Slovakia, Nr. 95/5195/134, \\
Ministry of Science and Technology of the Republic of Slovenia, \\
CICYT, Spain, AEN99-0950 and AEN99-0761,  \\
The Swedish Research Council,      \\
Particle Physics and Astronomy Research Council, UK, \\
Department of Energy, USA, DE-FG02-01ER41155, \\
EEC RTN contract HPRN-CT-00292-2002. \\

  


\end{document}